\documentclass[journal]{IEEEtran}

\usepackage[T1]{fontenc}
\usepackage{mathtools}
\usepackage{xurl}
\usepackage{amsmath}
\usepackage{amssymb}
\usepackage{amsthm}
\usepackage{amsfonts}
\usepackage{braket}

\usepackage{upgreek}
\usepackage{dirtytalk}

\usepackage{optidef}

\usepackage{algorithmic}
\usepackage{algorithm}
\usepackage{array}
\usepackage[caption=false,font=normalsize,labelfont=sf,textfont=sf]{subfig}
\usepackage{textcomp}
\usepackage{stfloats}
\usepackage{url}
\usepackage{verbatim}
\usepackage{graphicx}
\usepackage{setspace}
\usepackage{calc}
\usepackage{enumerate}
\usepackage[usenames]{color}

\usepackage{tabularx}
\usepackage{enumitem}
\usepackage{multirow}

\usepackage{amssymb}
\usepackage{upref}
\usepackage{epic,eepic}
\usepackage{times}
\usepackage{dsfont}
\usepackage{comment}
\usepackage{dirtytalk}
\usepackage{braket}
\usepackage{longtable}
\usepackage{hyperref}
\usepackage{xcolor}
\usepackage{xpatch}

\ifCLASSINFOpdf
\else
\fi
\usepackage{authblk}
\usepackage{cite}
\begin{document}

\title{Foundations of Quantum Federated Learning Over Classical and Quantum Networks \thanks{This work was supported in part through a Microgrant by Zaiku Group, the NSF Grants 2114267, CNS-1955744, NSF-ERC Center for Quantum Networks grant EEC-1941583, and MURI ARO Grant W911NF2110325.}
}
\author[1]{Mahdi Chehimi}
\author[2]{Samuel Yen-Chi Chen}
\author[1]{Walid Saad}
\author[3]{Don Towsley}
\author[4,5]{M\'erouane Debbah}
\affil[1]{\small Wireless@VT, Bradley Department of Electrical and Computer Engineering, Virginia Tech, Arlington, VA USA}
\affil[2]{\small Computational Science Initiative, Brookhaven National Laboratory, Upton, NY 11973, USA}
\affil[3]{\small University of Massachusetts Amherst, Amherst, MA USA}
\affil[4]{\small KU 6G Research Center, Khalifa University of Science and Technology, P O Box 127788, Abu Dhabi, UAE}
\affil[5]{\small CentraleSupelec, University Paris-Saclay, 91192 Gif-sur-Yvette, France}
\affil[ ]{\small \textit{\{mahdic,walids\}@vt.edu}, \textit{ychen@bnl.gov}, \textit{towsley@cs.umass.edu}, \textit{merouane.debbah@ku.ac.ae}\vspace{-0.5cm}}

\maketitle

\begin{abstract}
Quantum federated learning (QFL) is a novel framework that integrates the advantages of classical federated learning (FL) with the computational power of quantum technologies. This includes quantum computing and quantum machine learning (QML), enabling QFL to handle high-dimensional complex data. QFL can be deployed over both classical and quantum communication networks in order to benefit from information-theoretic security levels surpassing traditional FL frameworks. In this paper, we provide the first comprehensive investigation of the challenges and opportunities of QFL. We particularly examine the key components of QFL and identify the unique challenges that arise when deploying it over both classical and quantum networks. We then develop novel solutions and articulate promising research directions that can help address the identified challenges. We also provide actionable recommendations to advance the practical realization of QFL.
\end{abstract}

\IEEEpeerreviewmaketitle

\vspace{-0.17in}
\section{Introduction}\label{sec_intro}
\IEEEPARstart{F}{ederated} learning (FL) transformed the field of machine learning (ML) by promoting the shift from centralized, cloud-based learning to distributed, on-device edge learning. With FL, devices can collaborate in training local ML models by sending only their local model parameters to a central server. The server then aggregates the parameters, updates them, and sends the updated global parameters to all clients to repeat their local training. Using FL, edge devices maintain their local data, leading to increased privacy, communication efficiency, and scalability compared to centralized ML \cite{liu2022threats}.

State-of-the-art communication technologies and services incorporate large volumes of sensitive data about users' health, motion, activities, and social behavior that are used in training local ML models in FL frameworks. However, recent FL advances face various challenges that limit unleashing their full potential, which mainly include: 1) the increasing demand for stronger computational capabilities at edge devices due to growing data volumes and dimensionality, and 2) security and privacy risks stemming from attacks on the communication of FL learning parameters and vulnerabilities associated with untrusted servers \cite{liu2022threats}.

To address computational bottlenecks at the edge in FL, \emph{quantum federated learning (QFL)} was proposed \cite{chehimi2022quantumFL, chen2021federated_QML}. QFL can potentially leverage advancements in quantum computing for efficient, distributed quantum learning. In QFL, local clients utilize \emph{quantum machine learning (QML)} models, characterized by parametrized quantum circuits (PQCs) with classically-optimized parameters. These QML models can outperform their classical counterparts in complexity and computational efficiency, making QFL an effective solution for accelerating FL tasks.

The classically-optimized nature of QML parameters facilitates the initial deployment of QFL frameworks over existing classical communication infrastructures. Additionally, as \emph{quantum communication networks (QCNs)} mature, QFL is strategically positioned to leverage secure quantum communication protocols, like \emph{quantum key distribution (QKD)} and \emph{blind quantum computing (BQC)} to enhance the robustness of FL against parameter-centric attacks and mitigate server-based security vulnerabilities \cite{chehimi2022physics}. Implementing QFL over a classical network inherently presents numerous challenges. Furthermore, a seamless deployment of QFL over a QCN necessitates the incorporation of classical communication systems. This integration of both communication paradigms in QFL deployments intensifies existing challenges, which include QML scalability, hardware interface attenuation, and quantum noise considerations.

Prior works \cite{chehimi2022quantumFL,chen2021federated_QML,chehimi2022physics,zhao2023non,larasati2022quantum} studied the performance of QFL over classical networks, while utilizing both quantum \cite{chehimi2022quantumFL} and classical data \cite{chen2021federated_QML}. Furthermore, the works in \cite{chehimi2022quantumFL} and \cite{zhao2023non} studied the impact of data non-uniformity on overall QFL performance. However, each of these prior works \cite{chehimi2022quantumFL,chen2021federated_QML}, \cite{zhao2023non} focused solely on individual challenges confronting QFL, and none has provided a comprehensive analysis of QFL's challenges over classical networks. The authors in \cite{larasati2022quantum} provided a short analysis of QFL basics and challenges over classical networks. However, they did not provide an in-depth technical discussion of those challenges, nor did they shed light on potential solutions. Conversely, the examination of QFL over QCNs has been largely overlooked, even though existing literature has explored distributed learning and quantum computing in the context of QCNs \cite{chehimi2022physics,cacciapuoti2019quantum}, \cite{li2023entanglement}. To the best of our knowledge, there are no prior works that investigate the fundamentals of QFL, as well as the associated challenges and opportunities brought forward by deploying QFL over classical and quantum networks.

\begin{figure*}[t]
\begin{center}
\centerline{\includegraphics[width=2\columnwidth]{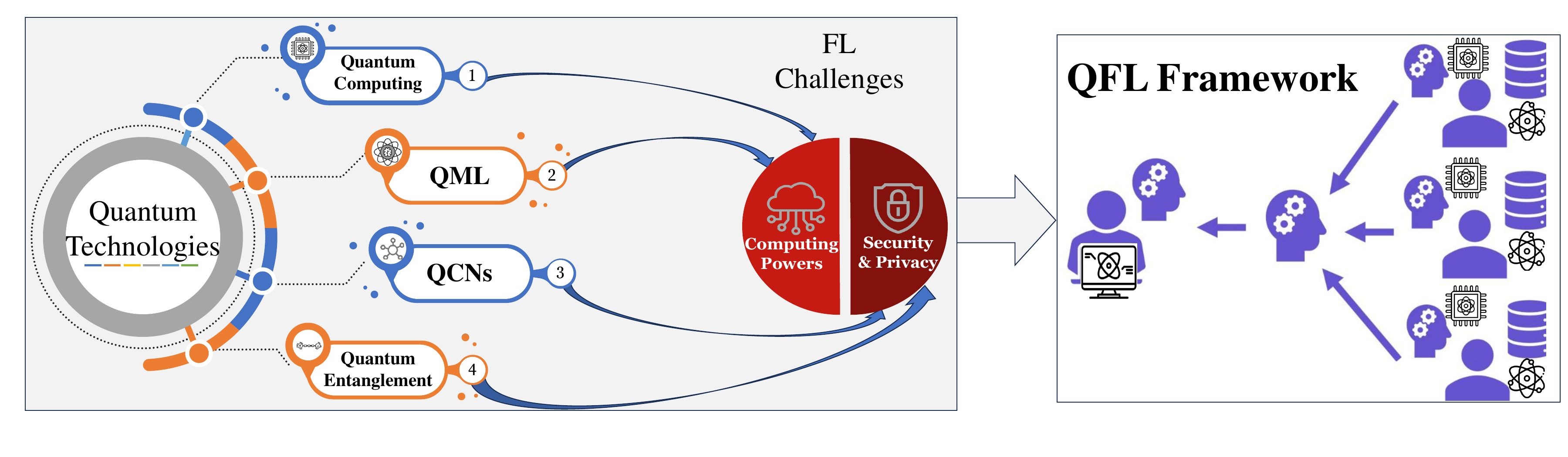}}\vspace{-0.6cm}
\caption{Proposed QFL framework at the intersection of FL and quantum technologies.}\vspace{-1cm}
\label{fig_contributions}
\end{center}
\end{figure*}

The main contribution of this work is the first thorough investigation of the integration of FL and different quantum technologies (see Fig.~\ref{fig_contributions}). In particular, we discuss the potential role of quantum computing, QML, and quantum communications in overcoming pressing FL challenges through the development of QFL. Our proposed framework lays down the foundations for deploying QFL over both classical and quantum networks. Towards this goal, we make the following key contributions:
\begin{itemize}
    \item We analyze potential opportunities related to the integration of QML and FL to create a QFL framework. We then study the opportunities and prospects of deploying the QFL framework over classical and quantum networks.

    \item We list and analyze the challenges facing practical deployments of QFL over classical networks. We thoroughly investigate the technical aspects of every challenge and its unique, quantum-specific impacts on QFL. Moreover, we propose several novel solutions and research directions that must be undertaken to overcome these challenges and ease the practical deployment of QFL over classical networks. 

    \item We present the first thorough investigation of the integration of QFL with QCNs. In particular, we articulate crucial research questions that need to be addressed and identify key challenges that hinder this integration. Moreover, we propose solutions and discuss fundamental research directions necessary to establish the foundations of QFL over QCNs.
\end{itemize}

\vspace{-0.1in}
\section{QFL over Classical Networks}
\subsection{Opportunities and Prospects}
Deploying QFL over classical networks offers a myriad of opportunities. In such a deployment, QFL can exploit quantum computers at the edge to run QML models, thus, providing enhanced capabilities for handling large-dimensional data. This is due to the intrinsic parallelism and proven efficiency of quantum computers \cite{abbas2021power}. Moreover, leveraging collaborative QML training across multiple quantum devices in QFL could offer faster and more efficient performance compared to centralized QML approaches. This is achieved through the aggregation of the classical QML learning parameters at a central server, which facilitates joint learning while preserving data privacy \cite{chehimi2022quantumFL}.

\vspace{-0.15in}
\subsection{Challenges}\label{sec_challenges_classical}
Real-world practical deployments of QFL over classical networks faces many key challenges, as discussed next.

\subsubsection{Quantum Gate Noise}
Local QML model training by QFL clients involves applying quantum gates and circuits, which, given the current state of \emph{noisy intermediate-scale quantum (NISQ)} devices, are highly susceptible to significant noise generation. This intrinsic local noise, produced during the training phase, will be inextricably integrated into the learning parameters, i.e., the quantum circuit parameters. Consequently, the information encapsulated within these learning parameters will contain a combination of useful local information and undesirable noise. This inherent embedding of noise within the learning parameters poses a substantial challenge in discerning the beneficial information from the noise \cite{cacciapuoti2019quantum}. Furthermore, at the QFL server, the aggregation of the intrinsically-noisy learning parameters from different clients may result in the obfuscation of useful information, leading to an extended training period and a consequent degradation in the overall QFL performance. One additional challenge here is the fact that the gate noise varies among different physical realizations of NISQ devices, which results in heterogeneous noise in the learning parameters at the server.

\subsubsection{Data Non-uniformity}
In QFL, various clients often possess \emph{non-independent and identically distributed (non-IID)} data, which follow distinct underlying distributions. As the number of clients with non-uniform and non-IID data participating in QFL increases, the necessary number of global communication rounds to accomplish training escalates accordingly \cite{zhao2023non}. While non-IID data is a well-known challenge for classical FL, this problem becomes more pronounced within the context of QFL when non-IID quantum data is leveraged (like quantum sensor data) due to the inherently-high dimensionality of such data and its unique properties, like superposition and entanglement \cite{chehimi2022quantumFL}. Accordingly, non-uniformity in the clients' data will pose significant \emph{communication overhead} during QFL training. Moreover, for efficient QFL performance, each client's data should be embedded into quantum states in a way that differences in the data are easily distinguished in the resulting Hilbert space. This allows for more accurate training and prediction. However, conventional quantum embedding techniques that rely on predefined quantum embedding circuits result in lossy training and degraded QFL performance. 

\begin{table*}[ht]
\caption{{Summary of challenges and proposed solutions for deploying QFL over a classical communication network.}}
\centering
{\renewcommand{\arraystretch}{0.9} 
\small 
\begin{tabularx}{\textwidth}{|c|X|}
\hline
\textbf{Challenges} & \textbf{Proposed Solutions}\\
\hline
\multirow{4}{*}{Quantum gate noise in QFL learning parameters \cite{cacciapuoti2019quantum}} & 
\begin{itemize}[leftmargin=*]
    \item New averaging algorithms with noise isolation at server \cite{yu2022quantum}.
    \item Probabilistic noise cancellation by QFL clients \cite{chehimi2022quantumFL}.
    \item Advanced QEC techniques \cite{cacciapuoti2019quantum}.
    \item Novel noise-capturing QFL performance metrics.
\end{itemize} \\
\hline
\multirow{2}{*}{Data non-uniformity and communication overhead \cite{zhao2023non}} & 
\begin{itemize}[leftmargin=*]
    \item Customized variational quantum data embedding circuits \cite{zhao2023non,chen2021federated_QML,chehimi2022quantum_semantic}.
    \item Real-world and synthetic non-uniform and non-IID federated quantum datasets \cite{chehimi2022quantumFL,zhao2023non}.
\end{itemize} \\
\hline
\multirow{2}{*}{Heterogeneous quantum capabilities \cite{chehimi2022physics}} & 
\begin{itemize}[leftmargin=*]
    \item QAS equipped with quantum search algorithms and QRL \cite{chen2021federated_QML}.
    \item Hardware-specific performance metrics to guide the transpilation operation \cite{cacciapuoti2019quantum}.
\end{itemize} \\
\hline
\multirow{3}{*}{Small-scale quantum devices \cite{chen2021federated_QML}} & 
\begin{itemize}[leftmargin=*]
    \item Advanced quantum hardware to have more qubits with longer coherence time \cite{li2023entanglement}.
    \item Data re-uploading techniques in local training of QML models \cite{chen2021federated_QML}.
    \item Quantum-inspired tensor networks for data compression \cite{chehimi2022quantum_semantic}.
\end{itemize} \\
\hline 
\multirow{3}{*}{Quantum-specific classical FL attacks \cite{liu2022threats}} & 
\begin{itemize}[leftmargin=*]
    \item Secure multi-party quantum computation protocols \cite{larasati2022quantum}.
    \item Quantum differential privacy.
    \item Blind quantum computing \cite{li2021quantum}.
\end{itemize} \\
\hline
\end{tabularx}}
\vspace{-0.2in}
\label{tab:summar_classical}
\end{table*}

\subsubsection{Heterogeneous Quantum Capabilities}
Participating QFL clients often exhibit heterogeneous quantum capabilities (numbers of available qubits, their fidelities, lifetimes, and memory decoherence characteristics) and utilize various qubit topologies (like superconducting qubits, and trapped ions), which introduce a unique set of challenges to QFL \cite{chehimi2022physics}. In general, QFL clients obtain similar high-level global learning parameters from a centralized server. While these global parameters correspond to a common QML circuit and quantum gate sequences, those gates and circuits are realized as distinct, hardware-specific instructions across diverse quantum computing platforms available to different QFL clients. In this regard, each QFL client must optimally \emph{transpile} the high-level global model onto its hardware topology. However, transpiling a quantum circuit from one topology to another often does not yield equivalent performance. In fact, it can commonly lead to degraded performance due to increased circuit depth or complexity. Moreover, every QFL client must optimize and map the global learning parameters to ones suited for its specific quantum hardware topology. After training, the parameters must be generalized again so that they can be shared with the server. This back-and-forth process can introduce additional noise and errors, which can degrade QFL performance.

\subsubsection{Small-scale Quantum Devices}
In the era of NISQ devices, the number of qubits available for manipulation within a quantum circuit is predominantly confined to the scale of hundreds of qubits \cite{chen2021federated_QML}. This constraint limits the amount of quantum gates and operations that can be executed and restricts the feasible depth of QML models. The number of currently feasible quantum gates and operations varies across different hardware realizations and is limited due to the inherent restrictions of current quantum technologies. Specifically, quantum circuit noise rapidly accumulates, and maintaining qubit coherence over extended periods is nontrivial \cite{chehimi2022physics}. The consequences of these limitations are particularly significant in QFL, since each client is constrained to train a small-scale local QML model. This inherently restricts the capacity to analyze high-dimensional data and substantially diminishes the potential benefits that would otherwise be gained from exploiting high-dimensional QML in QFL. As such, the small-scale nature of current NISQ devices poses a significant bottleneck to the expansion and maturation of QFL. 

\subsubsection{Quantum-specific Classical FL Attacks}
QML models in QFL frameworks introduce unique vulnerabilities, especially when deployed over classical networks. The direct transmission of QML classical learning parameters creates exposure to conventional FL attacks such as eavesdropping and parameter tampering. These vulnerabilities become more intricate due to quantum-specific features like quantum superposition which may be inherently encoded in the classical learning parameters. Key risks include \emph{membership inference attacks}, which, when integrated with quantum generative models, risk exposing sensitive data, and \emph{quantum shadow model attacks} that allow an adversary to approximate quantum circuit features \cite{liu2022threats}. Both elevate privacy and security concerns in QFL deployments over classical networks.

\vspace{-0.13in}
\subsection{Proposed Solutions and Future Directions}\vspace{-0.03in}
To address the aforementioned challenges, we next explore various research avenues and propose potential solutions, as summarized in Table \ref{tab:summar_classical}.

\subsubsection{Mitigating Noise and Errors} \label{sub_sub_noise}
In QFL, mitigating the noise present in the learning parameters entails the isolation of noise from the intrinsic useful information. This decoupling can be performed either by the server or client. At the server, new QFL averaging algorithms are needed to segregate noise from the beneficial client data distributions during the aggregation of client learning parameters, which is a nontrivial operation. Alternatively, at the client side, quantum error mitigation (QEM) techniques such as \emph{probabilistic noise cancellation} can be utilized. This approach involves numerous stochastic sampling operations, which can incur substantial computational overhead. While QEM techniques minimize noise and errors, quantum error correction (QEC) methods target completely removing errors, which makes them complex and computationally demanding as they require additional qubits for error detection and correction. Different quantum computing platforms vary in their QEC capabilities, impacting qubit fidelity during QML training which can affect the convergence of QFL. Advances in QEC are crucial for improving their efficiency and computational demands \cite{cacciapuoti2019quantum}. Finally, developing new performance metrics that quantify the intrinsic noise in QFL learning parameters and integrates the QML's quantum noise with conventional FL performance metrics is necessary to efficiently capture the QFL performance.

\subsubsection{Developing Variational Quantum Embedding Circuits}
In order to overcome the practical challenge of having non-uniform and non-IID client data in QFL, it is necessary to incorporate variational PQC architectures into the training process of local QML models. By doing so, each client can customize its embedding circuit to its unique data distribution. This will enhance the accuracy of QFL training while also minimizing the number of required training epochs. Moreover, to practically deploy QFL with non-uniform quantum data, there is a need to develop physics-based non-IID quantum federated datasets \cite{chehimi2022quantumFL}.

\subsubsection{Customizing QML Models and Transpilation}
To address the challenges due to the heterogeneity of the QFL client quantum resources, client QML model ansatzes must be optimally customized to their adopted technology, available resources, and input data distribution. To do so, a promising direction involves the application of \emph{quantum architecture search (QAS)} techniques, where the QML architecture that optimizes the overall QFL performance is selected. However, the search space for optimal QML architectures is vast, presenting a considerable challenge for conventional search algorithms. Here, \emph{quantum reinforcement learning (QRL)} and quantum search algorithms, e.g., Grover's search algorithm, offer promising solutions that can significantly accelerate the search process \cite{abbas2021power}. Furthermore, it is necessary to develop new performance metrics that capture the heterogeneity of quantum resources and guide the transpilation process for each quantum client. By incorporating such metrics in QFL and optimizing QAS and transpilation based on them, we can ensure that QML models are tailored to the specific capabilities and constraints of each client. Future research should further explore the practical steps needed to apply QAS techniques effectively.

\subsubsection{Optimizing Resource Utilization for Scalable QFL}
To address the limitations of NISQ devices and QML models, future work should prioritize the development of scalable QFL algorithms and protocols, as well as methods for efficient quantum computation distribution across multiple quantum processors \cite{larasati2022quantum}. Addressing this scalability challenge involves considerable hardware and software advances. From the hardware perspective, increasing the number of qubits and extending their coherence times in quantum computers is essential. Concurrently, on the software end, the development of advanced QEC techniques and new algorithms for effective utilization of available quantum resources is crucial. For example, methods that enable achieving more with fewer qubits, such as \emph{data re-uploading} models can be leveraged to train small-scale QML models over extremely high-dimensional datasets. In addition, quantum-inspired classical approaches, like tensor networks, may be a promising approach to enhance QFL scalability as they reduce the amount of processed data in QML models. 

\subsubsection{Enhancing Client Data Privacy}
The intersection of quantum cryptography, recognized for its capacity to facilitate unconditionally secure communication, and QFL, which is founded on the principle of privacy preservation amidst collaborative learning, presents a significant avenue for enhancing privacy in QFL. To thwart conventional FL attacks customized to QML models, quantum-based secure protocols over classical networks can be incorporated within the QFL framework to bolster client security and privacy. For instance, protocols such as secure multi-party quantum computation, which can help in calculating federated
gradients safely, could be adopted to secure sharing and aggregating QFL learning parameters \cite{yu2022quantum}. Additionally, quantum differential privacy can be leveraged to enhance the privacy of the learning parameters in QFL frameworks \cite{li2021quantum}. Such techniques will be integral to ensuring client privacy and security within QFL deployments over classical networks, thereby contributing to QFL's practical viability and effectiveness. Furthermore, techniques like BQC and QKD can be leveraged to secure QFL deployments over QCNs, as is discussed next. 

\vspace{-0.1in}
\section{QFL over Quantum Communication Networks}
\subsection{Opportunities and Prospects}
While existing distributed learning frameworks like classical FL and QFL over classical networks improve user privacy by not sharing client data over the network, their classical communication processes remain susceptible to security breaches that can leak learning parameters. In contrast, a QFL deployment over a QCN offers an extra layer of information-theoretic security as it allows the application of services such as QKD and BQC that are unique to a QCN. When QFL is deployed over a QCN, the transmission of learning parameters can be secured using entangled quantum states combined with QKD techniques. Such measures prevent the parameter leakage to adversaries, minimizing the risks associated with data reconstruction attacks. Moreover, by utilizing protocols like BQC, the joint learning in QFL can be secured even if the centralized server is not trustworthy. Lastly, unique quantum phenomena like \emph{quantum entanglement} inherent in a QCN has the potential to accelerate parameter sharing and enhance overall QFL performance \cite{li2023entanglement,chehimi2022physics}.

\begin{figure}[t]
\begin{center}
\centerline{\includegraphics[width=\columnwidth]{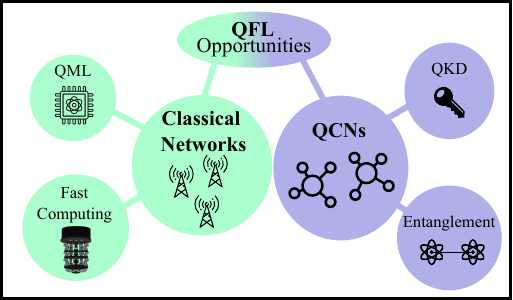}}\vspace{-0.1in}
\caption{Opportunities for QFL over classical networks and QCNs.}
\label{fig_opportunities}
\end{center}
\vskip -0.4in
\end{figure}

\vspace{-0.15in}
\subsection{Challenges}
We now turn our attention to challenges specific to deploying QFL frameworks over QCNs, where QML learning parameters are embedded and shared in quantum states and the QFL server has quantum capabilities. 

\subsubsection{Managing and Controlling Limited QCN Resources}
The performance of QFL over a QCN is constrained by the clients' available quantum resources and their characteristics, such as the quantity of successfully generated qubits, available quantum memory capacity, and coherent lifetime of qubits. Moreover, since storing quantum states in quantum memories entails decoherence, which degrades their quality or \emph{fidelity}, controlling and scheduling quantum memories is a challenging task that affects the quality of QFL learning parameters. For instance, quantum memories have limited capacities, and thus, some quantum states in memories must be discarded when stored for long durations, while newly generated quantum states must be stored at suitable memory locations to minimize their associated delays and gate errors, and maximize their fidelities \cite{chehimi2022physics}. In QFL, these characteristics will not only vary between different physical realizations, but will also vary between different clients with different capabilities, which further complicates the joint learning process. It is further challenging to create algorithms that provide accurate QFL performance and joint learning while controlling available QCN resources.

\subsubsection{Impact of Imperfect QCN Operations on QFL Accuracy}
When learning parameters are transferred as quantum information over a QCN, QCN nodes must perform multiple quantum operations, such as entanglement swapping and Bell-state measurements. However, due to practical imperfections, those operations often degrade the fidelity of the quantum information \cite{chehimi2022physics}. As such, a noisy version of the learning parameters will be received by the server, which can negatively affect the accuracy of QFL training, resulting in an increased number of communication rounds, thus imposing a communication overhead. The conventional approach in a QCN to enhance qubit fidelity relies on QEC or \emph{entanglement distillation}, where multiple low-fidelity qubits are consumed to generate fewer high-fidelity qubits \cite{li2023entanglement}. However, QFL expects various clients and QCN nodes to have different quantum capabilities, thus designing distillation protocols is challenging, because distillation itself can be lossy, and its performance varies with different quantum technologies and available resources.

\subsubsection{QFL De-synchronization and Training Latency} \label{sec_challenge_desynch}
Owing to the probabilistic characteristics inherent in various QCN operations, such as single-photon emissions and quantum measurements, success rates in such operations can differ among heterogeneous QFL clients. Consequently, the time required to prepare and transmit learning parameters via a QCN will vary across different clients after local training. This can result in \emph{de-synchronization} in the learning parameter aggregation process by the server, and hence, lead to increased \emph{QFL training latency}. For each QFL client, qubits generated at distinct time intervals are stored in quantum memories for varying durations, thereby experiencing different amounts of decoherence. When a qubit is transferred over a QCN from a client quantum memory towards the server, the coupling characteristics of the selected qubit and the technology in-use will result in different requisite times to execute quantum gates and operations on that specific qubit before being transferred. Additionally, QEC techniques, essential for QFL, involve encoding a smaller set of qubits using a larger qubit pool for error correction. However, the use of QEC requires the generation, distribution, and utilization of multiple qubits between remote nodes. Consequently, the iterative transfer of learning parameters over quantum channels may incur substantial delays due to QEC, decreasing the rate of QFL convergence. Furthermore, the classical communications required for QCN operations introduce latencies and de-synchronization in QFL training. These delays can adversely affect time-sensitive quantum operations, like Bell state measurements and entanglement swapping, thus acting as a bottleneck in QFL performance \cite{cacciapuoti2019quantum}.

\subsubsection{Heterogeneity in QCN and QFL Components}
Navigating the inherent technological discrepancies among various qubit types for a QFL framework deployed over a QCN presents a significant challenge. Specifically, qubits used in quantum computers for QML model execution and in quantum memories as \say{matter qubits} may differ from \say{flying qubits} used for parameter transmission and those used on the server-side. Each hardware technology excels in specific tasks but may be less suited for others, giving rise to a tradeoff between optimizing individual hardware components and maintaining technological consistency across the QFL framework for seamless operation \cite{chehimi2022physics}. A critical challenge here is the development of robust and precise interfaces that facilitate the transition of qubits between disparate quantum technologies, all while maintaining the integrity of the information they carry \cite{cacciapuoti2019quantum}.

\subsubsection{Impact of QCN Losses on QFL Generalizability}\label{sec_challenge_QFL_generalizability}
When QFL learning parameters are shared in qubits over quantum channels in QCNs, they are susceptible to several sources of loss, due to their fragile nature. In particular, when qubits, e.g., single photons of light, interact with their surrounding environment, they suffer from path-loss that scales exponentially with the travelled distance, and, accordingly, some qubits may be absorbed during this transmission \cite{chen2023q}. As a result, some learning parameters shared over QCNs may be lost during transmission, which can yield sparse learning parameters of some QFL clients during the aggregation step at the server. In such cases, QFL \emph{generalizability} (achieved when global parameter updates capture various non-uniform data distributions) will be degraded. This is because QFL training fails to capture the effect of clients' underlying data distributions with sparse transferred parameters, as these sparse parameters play a marginal role in the global parameter update step.

\vspace{-0.15in}
\subsection{Proposed Solutions and Future Directions}
Next, we propose solutions to the aforementioned challenges and identify research directions to enable QFL deployment over QCNs, as summarized in Table \ref{tab:summar_QCN}.

\begin{table*}[ht]
\caption{{Summary of challenges and proposed solutions for deploying QFL frameworks over QCNs.}}
\centering
{\renewcommand{\arraystretch}{0.9} 
\small 
\begin{tabularx}{\textwidth}{|c|X|}
\hline
\textbf{Challenges} & \textbf{Proposed Solutions} \\
\hline
\multirow{3}{*}{Limited QCN resources \cite{chehimi2022physics}} & 
\begin{itemize}[leftmargin=*]
    \item Novel quantum memory scheduling policies \cite{chen2023q}.
    \item QFL-specific resource allocation and association techniques for quantum states (online optimization and classical/quantum game theory) \cite{chehimi2023matching}.
    \item QFL aggregation algorithms incorporating quantum state fidelity \cite{li2021quantum}.
\end{itemize} \\
\hline
\multirow{3}{*}{Imperfect QCN Operations and distillation design \cite{chehimi2022physics,li2023entanglement}} & 
\begin{itemize}[leftmargin=*]
    \item Entanglement swap-distillation scheduling protocols \cite{chehimi2023scaling}.
    \item Joint QFL optimization framework for learning parameter aggregation and distillation design.
    \item Jointly considering number of distillation rounds and consumed states, states' fidelity, memory resources, QML size, and QFL accuracy \cite{chehimi2023scaling}. 
\end{itemize} \\
\hline
\multirow{4}{*}{De-synchronization and QFL training latency \cite{cacciapuoti2019quantum}} & 
\begin{itemize}[leftmargin=*]
    \item Incorporating quantum entanglement to transfer learning parameters \cite{li2023entanglement}.
    \item Redefine convergence bounds to incorporate probabilistic effects of quantum operations and quantum-specific training delays \cite{yu2022quantum}.
    \item QFI to develop lower bounds on QML performance in QFL, and speed-up QFL convergence while capturing entanglement impacts \cite{abbas2021power}.
    \item New performance metrics that consider fidelity and entanglement in QFL convergence analysis \cite{chehimi2022physics}.
\end{itemize} \\
\hline
\multirow{3}{*}{Heterogeneity in QCN and QFL components \cite{cacciapuoti2019quantum,chehimi2022physics}} & 
\begin{itemize}[leftmargin=*]
    \item Advanced transducer technology \cite{cacciapuoti2019quantum}.
    \item Novel physics-based algorithms to capture quantum state characteristics and fidelity over various technologies \cite{chehimi2022physics}.
    \item Adaptive QML training and QFL aggregating algorithms, incorporating various measurements of fidelity \cite{yu2022quantum}.
\end{itemize} \\
\hline 
\multirow{3}{*}{QCN losses and QFL generalizability \cite{chen2023q}} & 
\begin{itemize}[leftmargin=*]
    \item Quantum-compatible data compression techniques for resource efficient QCNs \cite{chehimi2022quantum_semantic}.
    \item Quantum network coding for reliable quantum communications \cite{li2023entanglement}.
    \item Semantic representations and quantum clustering for resilient learning parameters and generalizable QFL \cite{chehimi2022quantum_semantic}.
\end{itemize} \\
\hline
\end{tabularx}}
\vspace{-0.2in}
\label{tab:summar_QCN}
\end{table*}

\subsubsection{Efficient Control and Scheduling of QCN Resources} To overcome the QFL challenges associated with limited QCN resources, novel scheduling and resource allocation algorithms are needed. For instance, developing efficient quantum memory scheduling policies, which require careful analysis of cutoff times (times after which a stored qubit is discarded), is necessary to ensure a high quality of quantum states used to transfer QFL learning parameters \cite{chehimi2023matching}. Similarly, the optimal allocation of the generated quantum states to the available quantum memory is crucial for maximizing fidelity and, in turn, improving the quality of the transferred learning parameters and QFL accuracy. For this goal, tools from online optimization and \emph{classical and quantum game theory}, e.g., \emph{matching games}, can be utilized to achieve optimal associations between QCN resources. Finally, new QFL averaging algorithms are needed to incorporate the fidelities of quantum states that carry the learning parameters in the global QFL parameter update at the server side.

\subsubsection{Joint Optimization of QCN and QFL}
To enhance the fidelity of quantum states carrying QFL learning parameters over a QCN, novel entanglement swap-distillation scheduling algorithms are essential. These algorithms aim to maximize the fidelity of quantum states and thereby improve QFL accuracy. Concurrently, they must also ensure sufficient availability of QCN resources for the parameter transfer process. In addition, there is a need to develop a framework for the joint optimization of QFL learning parameter aggregation and QCN distillation operations in terms of the number of distillation rounds and the amount of quantum states consumed during distillation \cite{chehimi2023scaling}. This optimization framework must also account for available quantum memory resources, their fidelities, QML models' scale, and overall QFL accuracy \cite{chen2023q}.  

\subsubsection{QFL Convergence Analysis with Entanglement}
\emph{Quantum entanglement} can mitigate the QFL training de-synchronization challenge (see Section \ref{sec_challenge_desynch}). For instance, when QFL clients and server share entangled states, the direct transfer of learning parameters over quantum channels can be circumvented. Instead, protocols like entanglement teleportation and superdense coding can be employed, requiring only straightforward manipulation of the shared entangled states \cite{li2023entanglement}. Consequently, entanglement distribution can potentially minimize de-synchronization between learning parameters at the server, which helps to increase QFL convergence rates. Additionally, analysis of convergence rates for QFL over a QCN must be redone so as to capture quantum-specific training delays. For instance, concepts such as quantum Fisher information (QFI) should be incorporated to identify lower bounds on the training of local QML models. Moreover, QFI can be used to develop new algorithms to speed-up QFL convergence. Finally, new performance metrics that capture the fidelity of shared entangled states and their lifetimes must be integrated in the QFL convergence analysis \cite{abbas2021power}.

\subsubsection{Efficient Transducers and Adaptive QFL Algorithms}
To overcome the heterogeneity of QCN and QFL components as well as the different technologies used for each element of QFL over a QCN, novel \emph{transducer} hardware must be developed to efficiently map and transfer quantum states from one technology interface to another. Furthermore, novel algorithms with hardware-based controls and performance metrics must be developed to capture qubits characteristics and fidelities over the various quantum interfaces. Additionally, there is a need for adaptive QML training and QFL aggregation techniques that incorporate metrics and measures of qubit fidelity over all QFL elements \cite{cacciapuoti2019quantum}.

\subsubsection{Resource-efficient QCNs for Generalizable QFL}
To enhance QFL generalizability (see Section \ref{sec_challenge_QFL_generalizability}), non-uniform
client data distributions must be captured in the global learning
parameters. Henceforth, we must ensure reliable qubit transfer over QCNs to avoid losing information and producing sparse learning parameters. Accordingly, developing resource-efficient QCNs is necessary to minimize the amount
of resources utilized during the transfer of QFL learning parameters. This is because resource-efficient QCNs allow the use of higher quality QEC, which
ensures reliability in sharing learning parameters. Henceforth, quantum-compatible data compression must be investigated, since they can minimize the quantum data transmitted over QCNs, and enhance QFL efficiency. Additionally, \emph{quantum network coding} can enhance QCN efficiency and QFL generalizability. Finally, implementing dimensional reduction techniques, such as quantum clustering and semantic representations, is crucial before quantum data embedding in QFL to analyze hidden data structures. This process allows QML models to train using original data structure, enhancing resilience against parameter sparsity and improving QFL performance \cite{chehimi2022quantum_semantic}.


\section{Conclusion and Recommendations}\label{sec_conclusion}
QFL is an emerging field with significant practical potential, notably in sensitive sectors such as healthcare and military applications. In this paper, we have presented a comprehensive analysis of QFL deployments over both classical and quantum networks, while thoroughly examining the associated challenges and proposing solutions and future directions. 

Building on the proposed vision for QFL, we conclude with several recommendations, sequentially arranged from short-term to long-term priorities:

\subsubsection{Measuring Quantum Advantage}
Identifying when QFL outperforms classical FL requires advanced algorithms and benchmarks for accurate comparisons. Further, a detailed analysis of QML and QCN metrics is vital to establish QFL's advantages over such networks, aiding in pinpointing suitable practical use-cases.


\subsubsection{Classical-Quantum QFL Interoperability} To enable seamless QFL operation in hybrid quantum-classical network architectures, interfaces between quantum and classical optics must be developed. Optical fibers should also be optimized to handle both signal types, supported by efficient frequency conversion schemes in the semi-optical terahertz range. This will notably boost QFL's speed and efficiency.

\subsubsection{Ultimate QFL Security} While QCNs promise enhanced security for QFL, they are not immune to specific attacks and risks. Addressing these vulnerabilities calls for focused research in advanced quantum and \emph{post-quantum} cryptography, along with novel hardware-based designs of QFL's integration with BQC to fortify QFL's security infrastructure.

\vspace{-0.1in}
\bibliographystyle{IEEEtran}
\bibliography{references}




\end{document}